\documentclass[aps,notitlepage]{revtex4-1}

\usepackage{amssymb}
\usepackage{amsmath}
\usepackage{color}
\usepackage{cancel}
\usepackage{graphicx}
\usepackage{subfigure}
\usepackage[normalem]{ulem}
\newcommand{\be}{\begin{equation}}
\newcommand{\ee}{\end{equation}}
\newcommand{\ben}{\begin{eqnarray}}
\newcommand{\een}{\end{eqnarray}}

\newcommand{\DZero}{\rm D\O}

\begin{document}

\title{$X(3872)$ production and absorption in a hot hadron gas}

\author{L. M. Abreu$^{1}$, K. P. Khemchandani$^{2}$, A. Mart\'{i}nez Torres$^{3}$, 
F. S. Navarra$^{3}$, M. Nielsen$^{3}$ \\ 
}

\affiliation{$^{1}$Instituto de F{\'i}sica, Universidade Federal da Bahia, 40210-340, Salvador, BA, Brazil}

\affiliation{$^{2}$Faculdade de Tecnologia, Universidade do Estado do Rio de 
Janeiro, Rod. Presidente Dutra Km 298, P\'olo Industrial, 27537-000 , 
Resende, RJ, Brasil}

\affiliation{$^{3}$Instituto de F{\'i}sica, Universidade de S{\~a}o Paulo, Caixa Postal 66318, 05315-970, S{\~a}o Paulo, SP, Brazil}

\begin{abstract}

We calculate the time evolution of the  $X(3872)$ abundance in the hot hadron gas produced in the late stage of heavy ion collisions. We  use  effective field 
Lagrangians to obtain the production and dissociation cross sections of  $X(3872)$. In this evaluation we include diagrams involving the anomalous couplings 
$\pi D^*\bar{D}^*$ and $X \bar{D}^{\ast} D^{\ast}$ and also the couplings  of the $X(3872)$ with charged $D$ and $D^*$ mesons.  With these new terms the  $X(3872)$ 
interaction cross sections are much larger than those found in previous works.  Using these cross sections as input in rate equations, we conclude that during the 
expansion and cooling of the hadronic gas, the number of $X(3872)$, originally produced at the end of the mixed QGP/hadron gas phase, is reduced by a factor of 4.

\end{abstract}

\maketitle

\section{Introduction}

Over the last decade dozens of new charmonium states have been observed \cite{pdg,hosaka,esposito,nnl}. 
Among these new  states, the most studied one is  the $X(3872)$. It was first observed  in 2003 by  the Belle 
Collaboration \cite{Choi:2003ue,Adachi:2008te},  and has been confirmed by other five experiments: BaBar~\cite{Aubert:2008gu}, 
CDF~\cite{Acosta:2003zx,Abulencia:2006,Aaltonen:2009},  \DZero~\cite{Abazov:2004}, LHCb~\cite{Aaij:2011sn,Aaij:2013}
and  CMS~\cite{Chat13}. The LHCb collaboration has determined the 
$X(3872)$ quantum numbers to be $J^{PC} = 1^{++}$, with more than 8$\sigma$ 
significance \cite{Aaij:2013}.

The structure of the new charmonium states has been subject of an intense debate. In the case of  $X(3872)$, 
calculations using constituent quark models give masses for  possible charmonium states with $J^{PC}=1^{++}$ quantum 
numbers,  which are much bigger than the observed $X(3872)$ mass: $2~^3P_1(3990)$ and $3~^3P_1(4290)$ \cite{bg}. These results, 
together with the observed isospin violation in their hadronic decays, motivated the conjecture that these objects are multiquark states, 
such as mesonic molecules or tetraquarks. Indeed, the vicinity of the $X$ mass to the $\bar D D^{*}$  threshold inspired the proposal that 
the $X(3872)$ could be a molecular  $\bar D D^{*} $   bound 
state with a small binding energy \cite{swanson,eo}. Another well known interpretation of the  $X(3872)$ is that it can be a tetraquark state
resulting from the binding of a diquark and an antidiquark~\cite{maiani,ricardo}.  There are other proposals as well \cite{hosaka,esposito,nnl}. 
One successful approach in  describing experimental data is to treat the $X$ as an  admixture of two and four-quark states \cite{carina}. 

Until now it has not been possible to determine the strucute of the $X$, since the existing data on masses and decay widths can be explained by 
quite different models. However this situation can change, as we address 
the production of exotic charmonium in hadronic reactions, i.e. proton-proton,  proton-nucleus and nucleus-nucleus collisions. Hadronic collisions seem to be  
a promising testing ground for  ideas about the structure of the new states. It has been shown \cite{esposito,gri} that it is extremely difficult to 
produce meson molecules in p p collisions. In the molecular approach the estimated cross section for $X(3872)$ production is two orders of magnitude smaller 
than the measured one. On the other hand, in Ref. \cite{erike} a simple model was proposed to compute the $X$ production cross section in p p collisions in the 
tetraquark approach. Predictions were made for the  next run of the LHC.

As  pointed out in Refs.  \cite{Cho:2010db,EXHIC},  high energy heavy ion collisions offer an interesting 
scenario to study the production of multiquark states. In these processes, a quite large number of heavy quarks are expected to
be produced, reaching as much as 20 $c \, \bar{c}$ pairs per unit rapidity in Pb + Pb collisions at the LHC.  Moreover, the formation of quark gluon plasma 
(QGP) may enhance the production of  exotic charmonium states, since the charm quarks are free to move over
a  large volume and they may coalesce to form  bound states at the end of the QGP phase or, more precisely, at the end of the mixed phase, since the QGP needs some time 
to hadronize. 
One of the main conclusions of Refs.~\cite{Cho:2010db,EXHIC} was  that, if the production mechanism is coalescence, then  
the production yield of these exotic hadrons at the moment of their formation strongly reflects their internal structure. In particular  
it was shown   that in the coalescence model  the production yield of the $X(3872)$, at RHIC or LHC energies, is almost 20 times 
bigger  for a molecular structure  than for a tetraquark configuration.

After being produced at the end of the quark gluon plasma phase, 
the $X(3872)$  interacts with other hadrons during the expansion of the hadronic matter. Therefore, the $X(3872)$ can be  destroyed  in collisions 
with  the comoving  light  mesons, such as $X+\pi\to \bar D+D$, $X+\pi\to \bar D^* +D^*$,   but it can also
be produced through the inverse reactions, such as $D+\bar{D}\to X+\pi$, $\bar D^* +D^*\to \pi +X$.   We expect these cross sections to depend  
on the spatial configuration of the $X(3872)$.  Charm tetraquarks in a diquark - antidiquark configuration $(c q) -  (\bar{c}\bar{q})$ have a typical radius 
comparable to (or smaller than) the radius of the charmonium groundstates, i.e. $r_{4q} \simeq r_{\bar{c} c} \simeq 0.3 \, - \, 0.4 \,  \mbox{fm}$. 
Charm meson molecules are bound by meson exchange and  hence $r_{mol} \simeq 1/m_{\pi} \simeq 1.5$ fm. In fact, the calculations of Ref.  \cite{gamer} 
show that $r_{mol} \simeq  2.0 - 3.0 $ fm.  Molecules are thus much bigger than tetraquarks and their absorption cross sections may also 
be much bigger.  On the other hand, when these states are produced from  $D - \bar{D}^*$ fusion in a hadron gas,  what matters is the overlap between the 
initial and final state configurations. Assuming that the radius of the $D$ and $D^*$ mesons is $r_D \simeq   0.6 \,  \mbox{fm}$ \cite{haglin},  the 
initial $D \, + \,D^*$ state has a  larger spatial overlap with a molecule than with a tetraquark and, therefore, the production of molecules is favored.
Hence from geometrical arguments we expect that in a hot hadronic environment molecules are  easier to produce and also easier to destroy than tetraquarks.
Of course geometrical estimates of cross sections are more reliable if we apply them to high energy collisions. Here the typical collision energies are of 
the order of the temperature $T \simeq 100 - 180 $ MeV and are probably not high enough. Nevertheless, at a qualitative level, they can be useful as guidance.

In  Ref.~\cite{ChoLee} the interactions of the $X$ in a hadronic medium were studied in the framework of $SU(4)$ effective Lagrangians. The authors computed the 
corresponding production and absorption cross sections, finding that the absorption cross section is two orders of magnitude larger than the production one. 
The effective Lagrangians include the $X$ particle as a fundamental degree of freedom and the theory is unable to distinguish between molecular and tetraquark 
configurations. Presumably this information might be included in the form factors introduced in the vertices. 
The authors find that it is much easier to destroy the $X$ than to create it. In particular, for the largest thermally averaged cross 
sections they find:  $<\sigma v>_{\pi X  \rightarrow D^* \bar{D}^* } \,  \simeq 30 \, <\sigma v>_{D^* \bar{D}^* \rightarrow \pi X}$. In spite of this difference, 
the authors of  Ref. \cite{ChoLee}   arrived at the intriguing  conclusion that the number of $X$'s  stays approximately constant during the hadronic phase. In Ref. 
\cite{ChoLee} the coupling of the $X(3872)$ with charged charm mesons (such as $D^- \, D^{* +}$)  was neglected and only  neutral mesons were considered 
(such as $D^0 \bar{D}^{0*}$). Moreover, the terms with anomalous couplings  were not included in the calculations. 
In Ref.~\cite{XProduction} we  showed  that the inclusion of the couplings of the $X(3872)$ to charged $D$'s and $D^*$'s  and  those of the anomalous vertices, 
$\pi D^*\bar{D}^*$ and $X D^*\bar{D}^*$, 
increases the cross sections by more than one order of magnitude. Similar results were also observed in the case of $J/\psi\pi$ cross section \cite{nospsi}.
These anomalous vertices also give rise to new reaction channels, namely,  $\bar D+ D^*\to\pi +X$ and $\pi +X\to \bar D +D^*$.  Thus it is important to evaluate the 
changes that the above mentioned contributions can produce in the $X$ abundance (and in its time evolution) in reactions as those considered in Ref.  \cite{ChoLee}. 
This is the subject of this work.

The formalism used in Refs. \cite{ChoLee} and \cite{XProduction} was originally developed to study the interaction of charmonium states (specially the $J/\psi$) 
with light mesons in a hot hadron gas many years ago \cite{nospsi,rapp}.  The conclusions obtained in the past can help us now, giving some baseline for comparison. 
For example, if it is true that the $X$ has a large $c \bar{c}$ component, we may expect  that the corresponding  production and absorption cross sections are 
comparable to the $J/\psi$  ones.   
If, alternatively, they turn out to be much larger, this could be an indication of a strong multiquark and possibly molecular component.

The paper is organized as follows. In the next section we discuss the cross sections averaged over the thermal distributions.
In Sec.III we investigate the time evolution of the $X(3872)$  abundance  by  solving  the  kinetic  equation based on the phenomenological model of 
Ref. \cite{ChoLee}. Finally in Sec.IV we present our conclusions.

\section{Cross Sections averaged over the thermal distribution}

\begin{figure}[th]
\centering
\includegraphics[width=9.5cm]{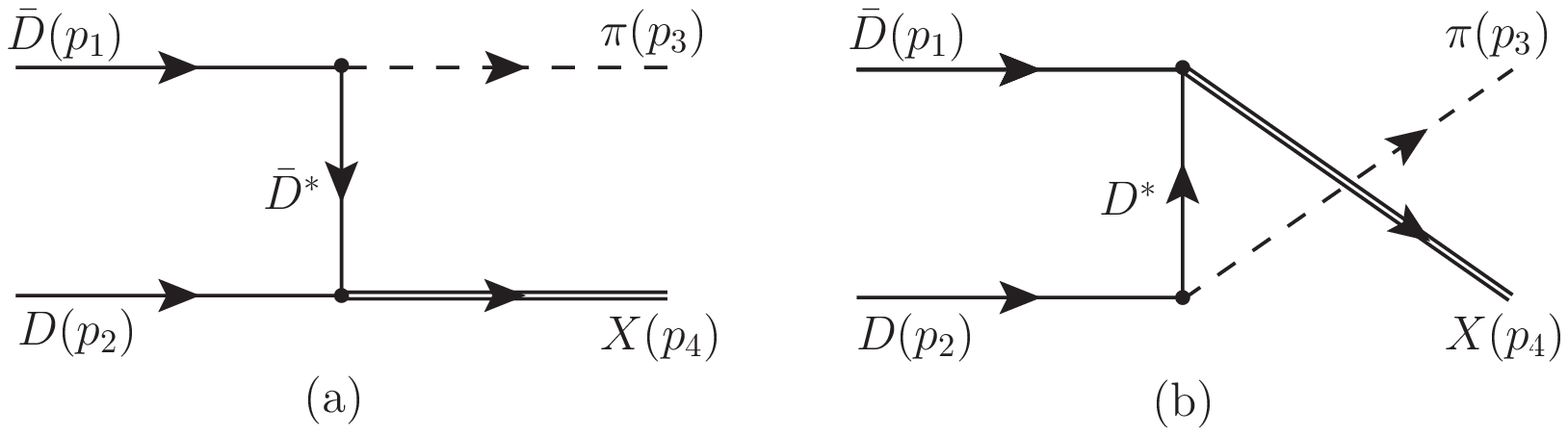}\\
\includegraphics[width=9.5cm]{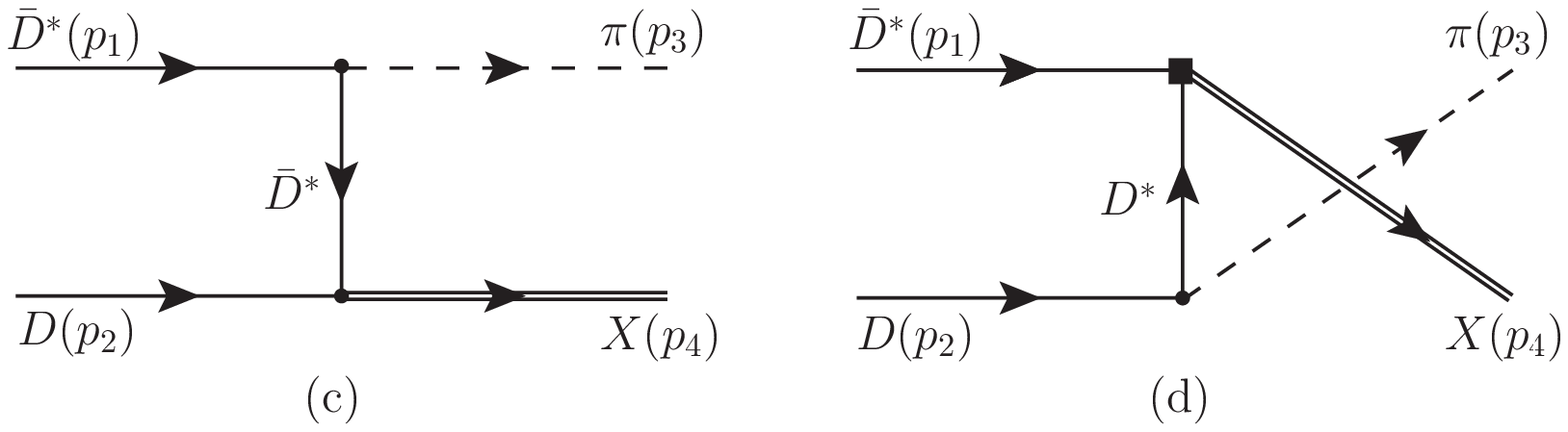}\\
\includegraphics[width=9.5cm]{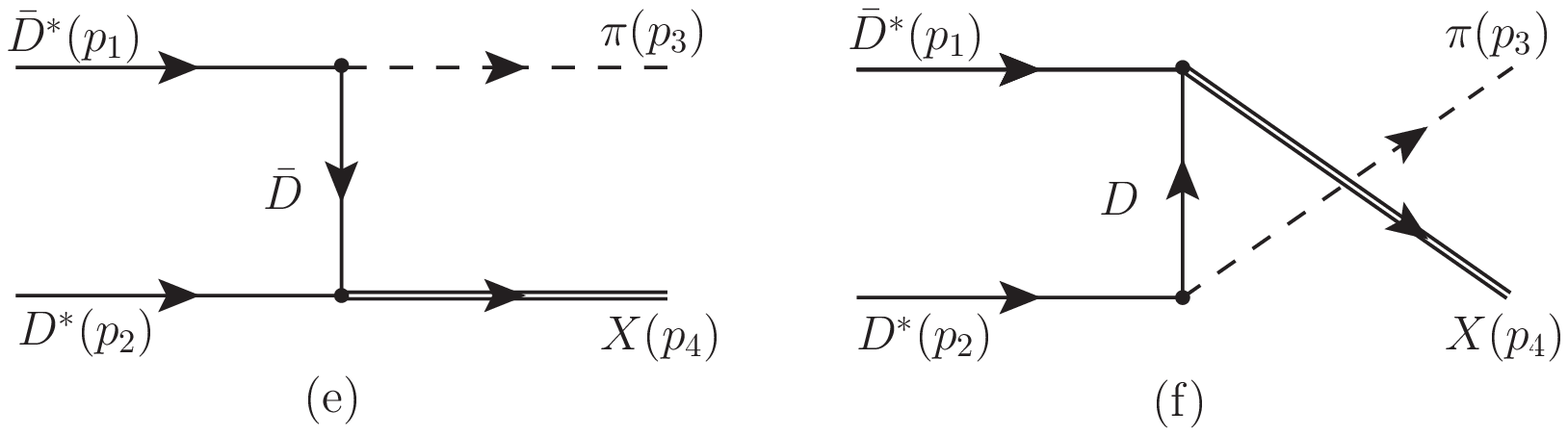} \\
\includegraphics[width=9.5cm]{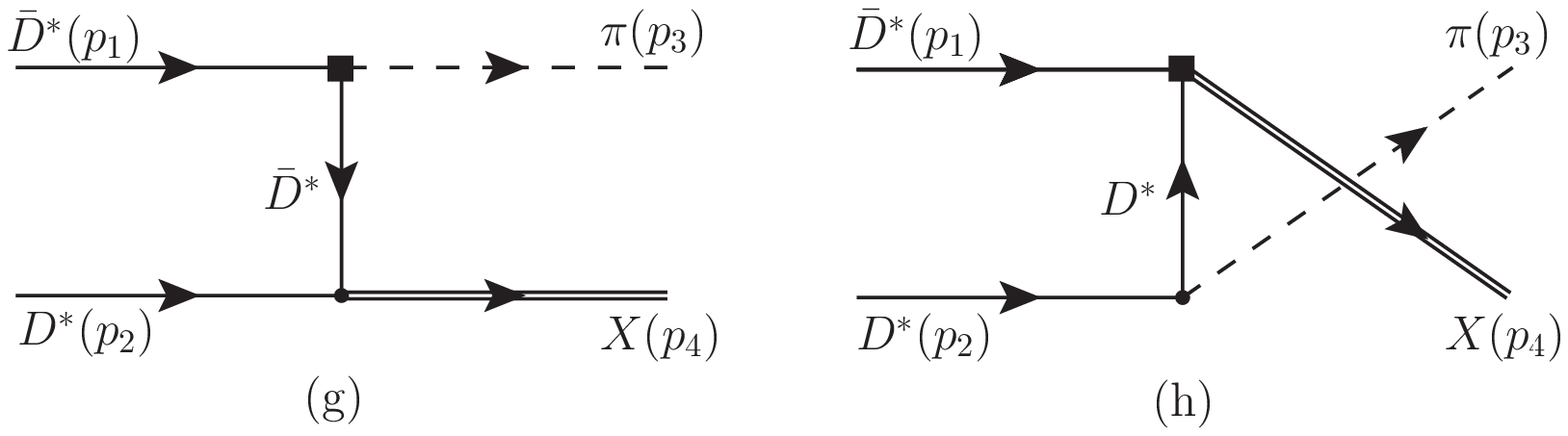} \\
\caption{Diagrams contributing to the processes $\bar{D} D \rightarrow \pi X$ [(a) and (b)], $\bar{D}^{\ast} D \rightarrow \pi X$ [(c) and (d)] 
and $\bar{D}^{\ast} D ^{\ast} \rightarrow \pi X $ [(e), (f), (g) and (h)]. The filled box in the diagrams (d), (g) and (h) represents the anomalous 
vertex $X D^*\bar{D}^*$, which was evaluated in Ref. \cite{XProduction}. The charges of the particles are not specified.}
\label{diagrams}
\end{figure}

In this section  we calculate the cross sections averaged over the thermal distributions for the processes $\bar D D\to \pi X$, $\bar D^* D\to \pi X$ 
and $\bar D^* D^*\to \pi X$, and for the inverse reactions. This information is the input to the study of  the time evolution of the $X(3872)$ abundance 
in hot hadronic matter. In Fig.~\ref{diagrams} we show the different diagrams contributing to each process. In Ref. \cite{ChoLee} it was shown that the contribution 
from the reactions involving the $\rho$ meson is very small compared to the reactions with pions and thus we  neglect the former in what follows.
The cross sections for the processes shown in Fig.~\ref{diagrams} were obtained in Ref.~\cite{XProduction}.  
Using effective Lagrangians based on SU(4), the coupling of the $X(3872)$ to  $\bar D^* D^*$  was 
estimated through the evaluation of triangular loops and an effective Lagrangian was proposed to describe this vertex. As a result, it 
was found that the contributions coming from the coupling of the $X(3872)$ to charged $D$'s and $D^*$'s and from the anomalous vertices play an 
important role in determining the cross sections. The coupling constant of the  $X\bar D^* D^*$ vertex  was found to be 
$g_{X\bar D^* D^*}=12.5 \pm 1.0$. For more details about the calculations, we refer the reader to Ref.~\cite{XProduction}. In the present 
manuscript, we follow Ref.~\cite{XProduction} and obtain 
the cross sections of the processes in Fig.~\ref{diagrams} using a form factor of the type
\be
F(\vec{q}) \, = \, \frac{\Lambda ^2}{\Lambda ^2 + \vec{q} ^2},
\label{formf}
\ee
where $\Lambda = 2.0$ GeV is the cutoff and $\vec{q}$ the three-momentum transfer in the center of mass frame.
Following Refs.~\cite{ChoLee,Koch},  the thermally averaged cross section for a process $a b \rightarrow c d$  can be calculated using the expression
\ben
\langle \sigma_{a b \rightarrow c d } \,  v_{a b}\rangle &  = & \frac{ \int  d^{3} \mathbf{p}_a  d^{3} \mathbf{p}_b \, f_a(\mathbf{p}_a) \, f_b(\mathbf{p}_b) \, 
\sigma_{a b \rightarrow c d } \,\,v_{a b} }{ \int  d^{3} \mathbf{p}_a  d^{3} \mathbf{p}_b \, f_a(\mathbf{p}_a) \,  f_b(\mathbf{p}_b) }
\nonumber \\
& = & \frac{1}{4 \alpha_a ^2 K_{2}(\alpha_a) \alpha_b ^2 K_{2}(\alpha_b) } \int _{z_0} ^{\infty } dz  K_{1}(z) \,\,\sigma (s=z^2 T^2) \left[ z^2 - (\alpha_a + \alpha_b)^2 \right]
\left[ z^2 - (\alpha_a - \alpha_b)^2 \right],
\label{thermavcs}
\een
where $f_a$ and $f_b$ are Bose-Einstein  distributions, $\sigma_{a b \rightarrow c d }$ are the cross sections computed in \cite{XProduction},   
$v_{ab}$ represents the relative velocity of the  two  interacting particles ($a$ and $b$), $\alpha _i = m_i / T$, with $m_i$ being the mass of 
particle $i$ and $T$ the temperature, $z_0 = max(\alpha_a + \alpha_b,\alpha_c + \alpha_d)$, and $K_1$ and $K_2$  are the modified Bessel functions of 
first and second kind, respectively.  The  masses used in the present work are: $m_D = 1867.2 $ MeV, $m_{D^{\ast}} = 2008.6$ MeV, 
$m_{\pi} = 137.3$ MeV and $m_{X} = 3871.7$ MeV~\cite{pdg}.

\begin{figure}[ht!]
\begin{center}
\subfigure[ ]{\label{fig1a}
\includegraphics[width=0.32\textwidth]{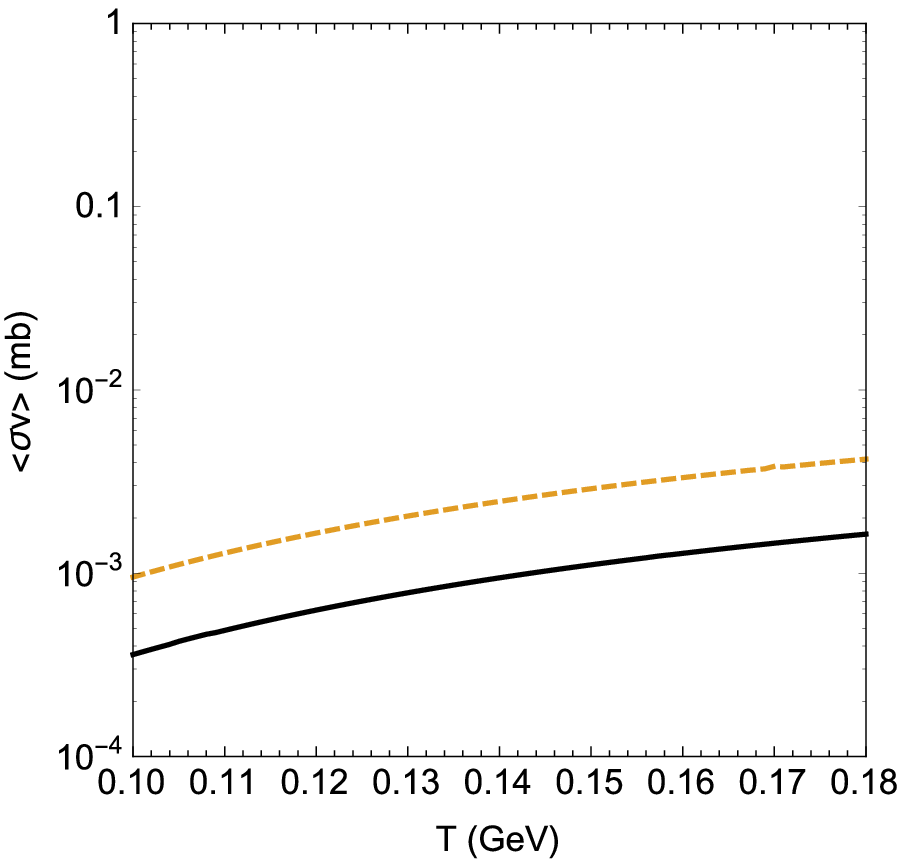}}
\subfigure[ ]{\label{fig1b}
\includegraphics[width=0.32\textwidth]{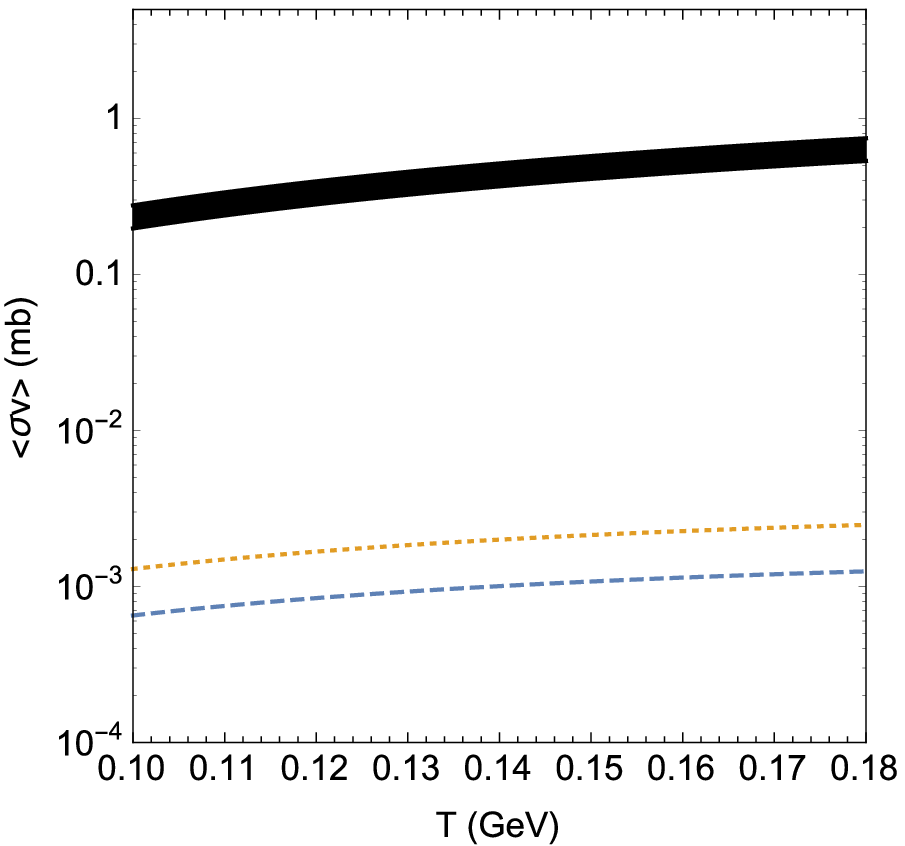}}
\subfigure[ ]{\label{fig1b}
\includegraphics[width=0.32\textwidth]{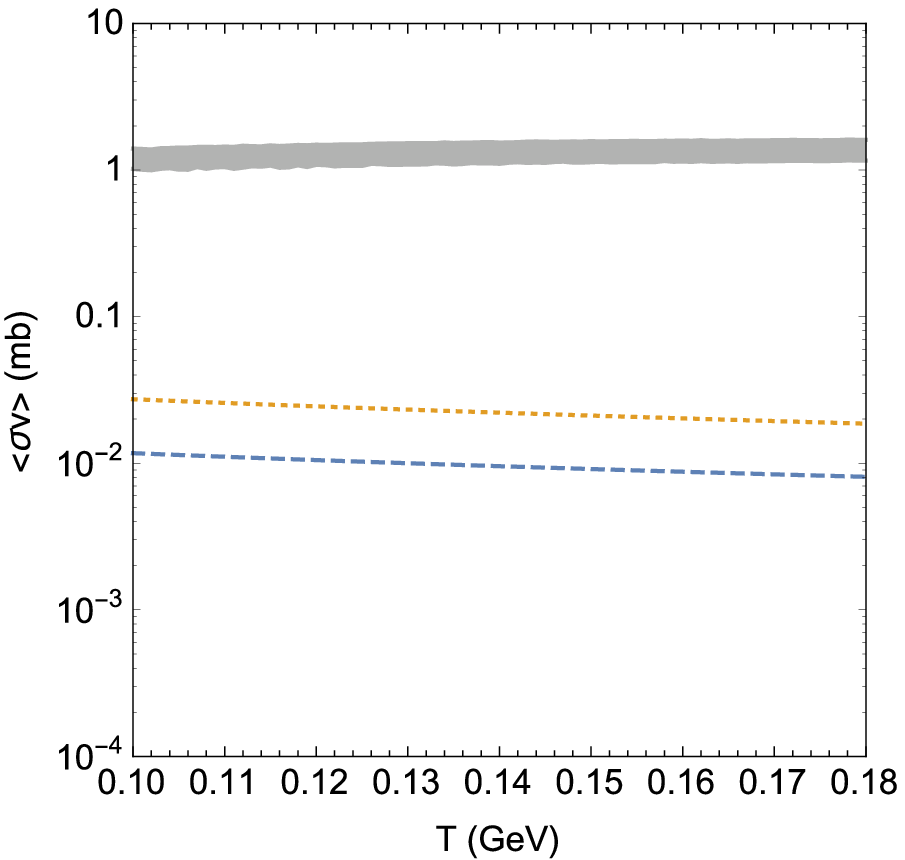}}
\end{center}
\caption{Thermally averaged cross sections. a)  $\bar{D} D \rightarrow \pi X(3872) $, considering only the coupling of the $X$ to the  neutral 
$D$'s and $D^*$'s  (solid line) and adding the coupling to charged $D$'s and $D^*$'s (dashed line).
b)  $\bar{D}^{\ast} D \rightarrow \pi X(3872) $, considering  only the coupling of the 
$X$ to  neutral $D$'s and $D^*$'s  (dashed line),  including the contribution from  charged $D$'s and $D^*$'s (dotted line) 
and also including diagrams containing the anomalous vertices (shaded region). 
c)  $\bar{D}^{\ast} D^{\ast} \rightarrow \pi X(3872)$. The lines and shaded area have the same meaning as those of b).
}
\label{fig2}
\end{figure}

In Fig. \ref{fig2}a  we show the thermally averaged cross section for the process $\bar{D} D \rightarrow \pi X(3872) $, considering only the  coupling of 
the $X(3872)$ to the neutral states  $\bar D^0D^{*0}$ (solid line) and adding the coupling to the charged components (dashed line). As  can be seen, 
the thermally averaged cross section increases by a factor of about 2.5 when we  include the charged $D$'s and $D^*$'s.

In Figs. \ref{fig2}b and \ref{fig2}c  we show the thermally averaged cross sections for the processes $\bar{D}^{\ast} D \rightarrow \pi X(3872) $ 
and  $\bar{D}^{\ast} D^{\ast} \rightarrow \pi X(3872) $,  considering only the coupling of the $X$ to  neutral $D$'s and $D^*$'s (dashed line), 
including couplings to charged  $D$'s and $D^*$'s (dotted line) and  finally adding also  the contribution from the anomalous vertices (shadded region).  
As can be seen,  the contribution from the anomalous vertices  produces  an enhancement of  the thermally averaged cross sections by 
a factor of  $100 \, - \, 150$.

To close this section we show  in Fig. \ref{fig3}a    the total thermally averaged cross sections for the {processes involving the production of the 
$X(3872)$ state, i.e., $\bar{D} D \rightarrow \pi X$, $\bar{D}^{\ast} D \rightarrow \pi X$ and $\bar{D}^{\ast} D ^{\ast} \rightarrow \pi X $ reactions, 
while in Fig.~\ref{fig3}b we show the inverse processes, i.e., the dissociation of $X(3872)$ through the reactions 
$\pi X\to \bar D D$, $\pi X\to \bar D^* D$, $\pi X\to \bar D^* D^*$, respectively.  For the latter cases, we  use the principle of 
detailed balance to determine the corresponding cross sections.  Figure \ref{fig3} should be compared with the Fig. 3 of Ref. \cite{ChoLee}. Our cross 
sections are a factor $100$ larger in reactions with $D^* \bar{D}^*$ in the initial or final state  This can be attributed mostly to the inclusion of the 
anomalous terms. Moreover, our cross sections are a factor $10$ larger in the case of $D D$ mesons in the initial or final state. The difference comes from the 
inclusion of the coupling of the $X$ to charged charged $D$'s and $D^*$'s, which was not considered in Ref. \cite{ChoLee}.

On the other hand, in both works, the absorption cross sections are more than fifty times larger than the production ones.

In the computation of the time evolution of the  $X$ abundance, we will need to know how the temperature changes with time and this is 
highly model dependent. Fortunately, as one can see in  Fig. \ref{fig3}, the dependence of $<\sigma v>$ on the temperature is relatively weak.

\begin{figure}[ht!]
\begin{center}
\subfigure[ ]{\label{fig1a}
\includegraphics[width=0.485\textwidth]{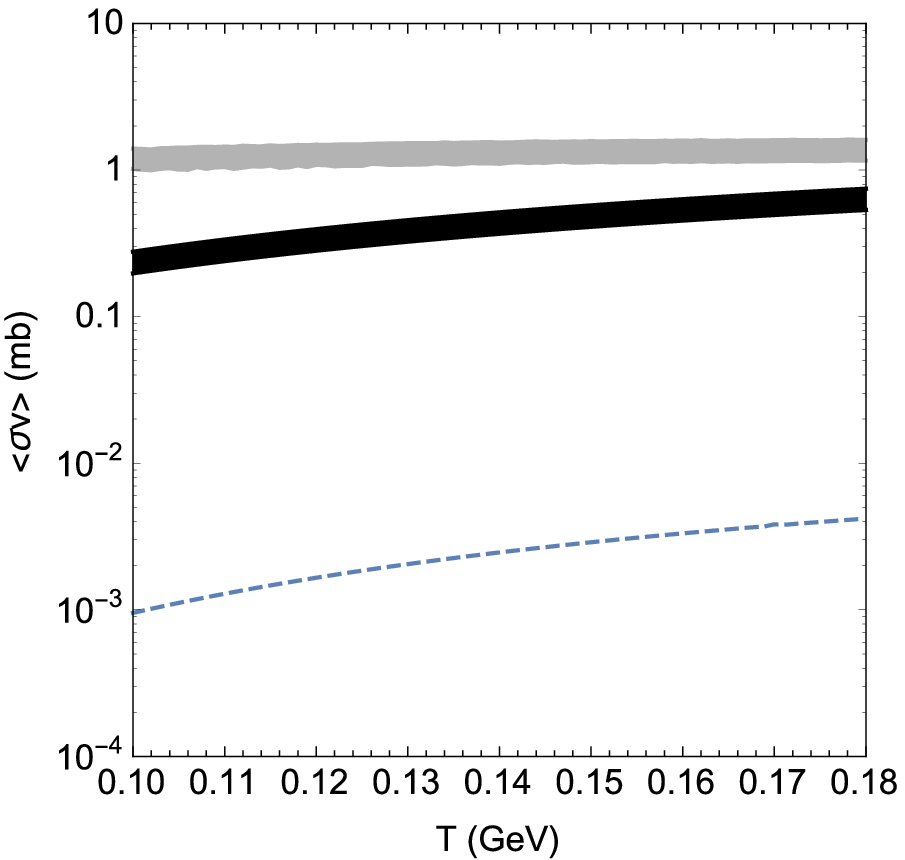}}
\subfigure[ ]{\label{fig1b}
\includegraphics[width=0.485\textwidth]{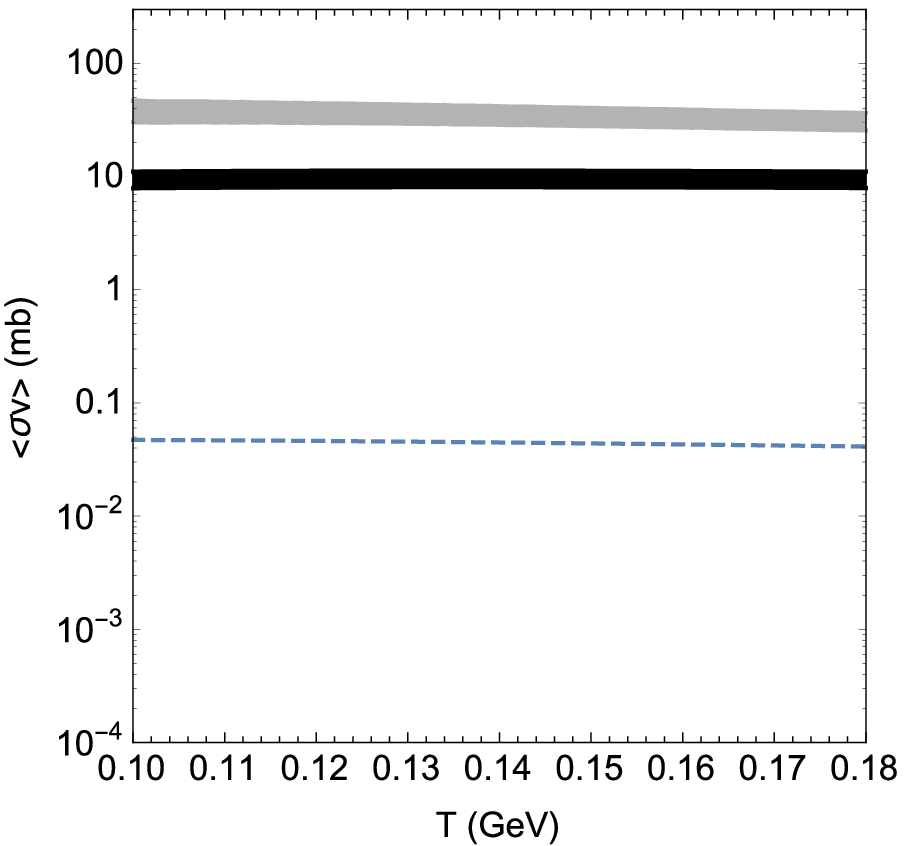}}
\end{center}
\caption{Thermally averaged cross sections. 
a) $\bar{D} D \rightarrow \pi X(3872) $ (dashed line), $\bar{D}^{\ast} D \rightarrow \pi X(3872) $ (dark-shaded region) and 
$\bar{D}^{\ast} D^{\ast} \rightarrow \pi X(3872) $ (light-shaded region). 
b) $\pi X(3872) \rightarrow \bar{D} D $ (dashed line), $\pi X(3872) \rightarrow \bar{D}^{\ast} D  $ (dark-shaded region) and 
$\pi X(3872) \rightarrow \bar{D}^{\ast} D^{\ast}$ (light-shaded region).}
\label{fig3}
\end{figure}

\section{Time evolution of the $X(3872)$ abundance}

Following Ref. \cite{ChoLee} we  study the yield of $X(3872)$ in  central Au-Au collisions at $\sqrt{s_{NN}} = 200$ GeV. 
By using the thermally averaged cross sections  obtained in the previous section,  we can now analyze the time evolution of the $X(3872)$ 
abundance in hadronic matter, which depends on the densities and abundances of  the particles involved  in  the processes of Fig.~\ref{diagrams},  
as well as the cross sections associated with these reactions (and the corresponding inverse reactions), Figs.~\ref{fig3}a  and 
\ref{fig3}b. The momentum-integrated evolution equation has the form~\cite{ChoLee,Koch,ChenPLB,ChenPRC}
\ben
\frac{d N_{X} (\tau)}{d \tau} & = & R_{QGP} (\tau) + \sum_{c,c^{\prime}} \left[ \langle \sigma_{c c^{\prime} \rightarrow \pi X } 
v_{c c^{\prime}} \rangle n_{c} (\tau) N_{c^{\prime}}(\tau)
- \langle \sigma_{ \pi X \rightarrow c c^{\prime} } v_{ \pi X} \rangle n_{\pi} (\tau) N_{X}(\tau)  \right], 
\label{rateeq}
\een
where $N_{X} (\tau)$, $N_{c^{\prime}}(\tau)$,  $ n_{c} (\tau)$ and $n_{\pi} (\tau)$ are the abundances of $X(3872)$, of  charmed 
mesons of type $c^{\prime}$, of  charmed mesons of type $c$ and of  pions  at proper time $\tau$, respectively. The term $R_{QGP} (\tau)$   
in Eq.~(\ref{rateeq})  represents the $X$ production from the quark-gluon plasma in the mixed phase, since the hadronization 
of the QGP takes a finite time,  and it is given by~\cite{ChoLee,ChenPRC}: 
\begin{align}
R_{QGP} (\tau)=\left\{\begin{array}{c} \dfrac{N^0_X}{\tau_H-\tau_C},\quad \tau_C<\tau<\tau_H,\\\\ 0,\quad\text{otherwise, }\end{array}\right.\label{R} 
\end{align}
where the times $\tau_C = 5.0$ fm/c and $\tau_H = 7.5$ fm/c determine the beginning and the end of the  mixed phase respectively.    
The constant $N^0_X$ corresponds to the total number of $X(3872)$ produced from quark-gluon plasma.}
To solve Eq.~(\ref{rateeq}) we assume that  the pions and charmed mesons  in the reactions contributing to the abundance of $X$ are in equilibrium. 
Therefore $N_{c^{\prime}}(\tau)$, $ n_{c} (\tau)$ and $n_{\pi} (\tau)$ can be written as~\cite{ChoLee,Koch,ChenPLB,ChenPRC}
\ben
N_{c^{\prime}}(\tau) & \approx & \frac{1}{2 \pi^2} \, \gamma_{C} \,  g_{D} \,  m_{D^{(\ast)}}^2  \, T(\tau) \,  V(\tau) \, 
K_{2}\left(\frac{m_{D^{(\ast)}} }{T(\tau)}\right) ,\nonumber \\
n_{c} (\tau) &  \approx & \frac{1}{2 \pi^2} \, \gamma_{C} \,  g_{D} \,  m_{D^{(\ast)}}^2 \,  T(\tau) \, 
K_{2}\left(\frac{m_{D^{(\ast)}} }{T(\tau)}\right), \nonumber \\
n_{\pi} (\tau) &  \approx & \frac{1}{2 \pi^2} \, \gamma_{\pi} \,  g_{\pi} \, m_{\pi}^2 \,  T(\tau) \, 
K_{2}\left(\frac{m_{\pi} }{T(\tau)}\right), 
\label{densities}
\een
where $\gamma _i$ and $g_i$ are the fugacity factor and the spin degeneracy of  particle $i$ respectively. As  can be seen in Eq.~(\ref{densities}), 
the time dependence in Eq.~(\ref{rateeq}) enters through the parametrization of the temperature $T(\tau)$ and volume $V(\tau)$ profiles suitable to 
describe the dynamics of the hot hadron gas after the end of the quark-gluon plasma phase.  
Following Refs.~\cite{ChoLee,ChenPLB,ChenPRC},  we assume the $\tau$ dependence of $T$ and $V$ to be given by~\cite{ChoLee,ChenPLB,ChenPRC} 
\ben
T(\tau) & = & T_C - \left( T_H - T_F \right) \left( \frac{\tau - \tau _H }{\tau _F - \tau _H}\right)^{\frac{4}{5}} , \nonumber \\
V(\tau) & = & \pi \left[ R_C + v_C \left(\tau - \tau_C\right) + \frac{a_C}{2} \left(\tau - \tau_C\right)^2 \right]^2 \tau_C .
\label{TempVol}
\een
These expressions are based on the boost invariant picture of Bjorken~\cite{Bjorken} with an accelerated transverse expansion.  In the above equation 
$R_C = 8.0$ fm denotes the final  size of the quark-gluon plasma, while $v_C = 0.4 \, c$  and  $a_C = 0.02 \,  c^2$/fm are its transverse flow velocity 
and transverse acceleration at $\tau_C$ respectively.  The critical temperature of the  quark gluon plasma to hadronic matter transition  is $T_C=175$ MeV; 
$T_H = T_C = 175$ MeV is the temperature of the hadronic matter at the end of the mixed phase. The freeze-out takes place at the freeze-out time $\tau_F = 17.3$ fm/c, 
when the temperature drops to  $T_F = 125$ MeV.

To solve Eq.~(\ref{rateeq}), we assume that the total number of charm quarks in charm hadrons is conserved during the  production and 
dissociation reactions, and that the total number of charm quark pairs  produced at the initial stage of the collisions at RHIC is 3, yielding the charm 
quark fugacity factor $\gamma _C \approx 6.4$ in Eq.~(\ref{densities}) \cite{ChoLee,EXHIC}. In the case of pions, we follow Ref.~\cite{ChenPRC} 
and work with the assumption that their total number at freeze-out is 926, which fixes the value of $\gamma _{\pi}$ appearing in Eq.~(\ref{densities}) 
to be $\sim 1.725$ . 

In Ref. \cite{ChoLee} the authors studied the yields obtained for the $X(3872)$ abundance within 
two different approaches: the statistical and the coalescence models. In the statistical model, hadrons are produced in thermal and chemical equilibrium. 
This model does not contain any information related to the internal structure of the $X(3872)$ and, for this reason 
we do not consider it in this work. 
In the case of the coalescence model, the determination of the abundance of a certain hadron is based on
the overlap of the density matrix of the constituents in an emission source with the Wigner function of the produced
particle.  This model contains information   on the internal structure of the considered hadron, such as 
angular momentum, multiplicity of quarks, etc. According to Ref.~\cite{EXHIC}, the number of $X(3872)$ 
produced  at the end of the mixed phase, assuming that the $X(3872)$ is a tetraquark state with
$J^{PC}=1^{++}$, is given by:
\begin{align}
N_{X(4q)} ^{0}=N_{X(4q)}(\tau_H) \approx  4.0 \times 10^{-5} .  \label{Nx4q}
\end{align}

\begin{figure}[th]
\centering
\includegraphics[width=8.0cm]{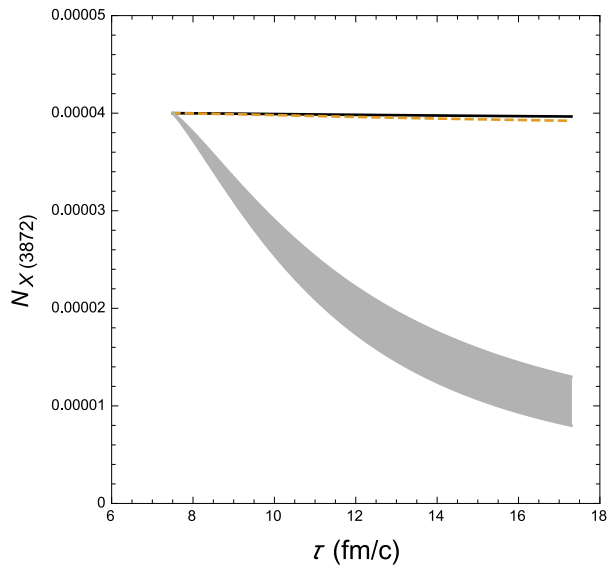}
\caption{Time evolution of the $X(3872)$ abundance as a function of  the proper time $\tau$ in central Au-Au collisions at $\sqrt{s_{NN}} = 200$ GeV. 
The solid line, the  dashed line and the light-shaded region represent the  results obtained considering only  the neutral $D$'s and $D^*$'s, 
adding the contribution from charged $D$'s and $D^*$'s  and including contributions from the anomalous vertices respectively. The initial condition 
is the abundance of the $X(3872)$ considered as a tetraquark Eq. (\ref{Nx4q}).}
\label{fig4}
\end{figure}

In order to determine the time evolution of the $X(3872)$ abundance we solve Eq.~(\ref{rateeq}) starting at the end of the mixed phase, i.e. at 
$\tau_H = 7.5$ fm/c,  and assuming that the $X(3872)$ is a tetraquark, formed according to the prescription of the coalescence model. The initial condition is 
given by Eq.~(\ref{Nx4q}). 
We use this initial abundance to integrate  Eq. (\ref{rateeq}) and we show the result in Fig.~\ref{fig4}. 
In the figure the solid line represents the result obtained using the same approximations as those made  
in Ref. \cite{ChoLee}. Our curve is slightly different from that of Ref.  \cite{ChoLee} because we did not include the  contribution of the 
$\rho$ mesons, as discussed earlier.  The dashed line shows the result when we include the couplings of the $X(3872)$ to charged $D$'s and $D^*$'s. 
The light-shaded band 
shows the results obtained with the further inclusion of the diagrams containing the anomalous vertices. The band reflects the uncertainty in the 
$X \bar{D}^* D^*$ coupling constant, which is $g_{X \bar{D}^{\ast} D^{\ast}} =  12.5\pm 1.0$ \cite{XProduction}.     

As  can be seen, without the inclusion of the anomalous coupling terms, the abundance of $X$ remains basically constant. 
This is because  the magnitude of the thermally averaged cross sections for the $X$ production and absorption reactions obtained within this 
approximation is so small that the second term in the right hand side of Eq.~(\ref{rateeq})
is completely negligible compared to the first term. 
When including the coupling of the $X$ to  charged $D$'s and $D^*$'s  we basically do not find any important change 
for the time evolution  of the $X$ abundance, since, as  can be seen in Fig.~\ref{fig2}, the thermally averaged cross sections do not change 
drastically in this case. On the other hand, the  inclusion of the anomalous coupling terms, depicted in Figs.~\ref{diagrams}c, \ref{diagrams}d, 
\ref{diagrams}f, \ref{diagrams}g and \ref{diagrams}h, 
modifies the behavior of the $X(3872)$ yield,  producing a fast decrease 
of the $X$ abundance with time. This is the main result of this work. We emphasize that the $X(3872)$ abundance, whose  time evolution was studied above, 
 is the only 
one which comes from the QGP and is what could be observed if the $X(3872)$ is a tetraquark state.
However, if the $X(3872)$ is a molecular state, it will be formed by hadron coalescence at the end of the hadronic phase. According to Ref.~\cite{ChoLee}, 
at this time the average number of $X$'s, considering it as a $D\bar{D}^*$ molecule, is 
\begin{align}
N_{X(mol)} \approx  7.8 \times 10^{-4}\;, \label{Nxmol}
\end{align}
which is about $80$ times larger than the contribution for a tetraquark state at the end of the hadronic
phase (see Fig.~\ref{fig4}). We can then conclude that 
the QGP contribution for the $X(3872)$ production (as a tetraquark state and from quark coalescence) after 
being suppressed during the hadronic phase, becomes insignificant at the end of the hadronic phase.

\section{Concluding Remarks}

In this work we have studied the time evolution of the  $X(3872)$ abundance in heavy ion collisions. 
If the $X(3872)$ is a tetraquark state it will be produced at the  mixed phase by quark coalescence.
After being produced at the end of the quark gluon plasma phase, 
the $X(3872)$  interacts with other comoving hadrons during the expansion of the hadronic matter. Therefore, the $X(3872)$ can be  destroyed  in collisions 
with  the comoving  light  mesons, such as $X+\pi\to \bar D+D$, $X+\pi\to \bar D^* +D^*$  but it can also
be produced through the inverse reactions, such as $D+\bar{D}\to X+\pi$, $\bar D^* +D^*\to \pi +X$.
In this work we have considered the contributions of anomalous  vertices, $\pi D^*\bar{D}^*$ and $X \bar{D}^{\ast} D^{\ast}$, and the contributions 
from charged  $D$ and $D^*$ mesons   to  the $X(3872)$ 
production and dissociation cross sections. These vertices, apart from enhancing the cross sections associated with the $\bar D^* D^*$ channel,   give 
rise to additional production/absorption mechanisms of $X$, which are found to be relevant.

The cross sections, averaged over the thermal distribution, have been used to analyze the time evolution of the $X(3872)$ abundance in hadronic matter. 
We have found that the abundance of a tetraquark
$X$ drops from $N_{X(4q)} \approx  4.0 \times 10^{-5}$ at the begining of the hadronic phase
\cite{ChoLee} to $N_{X(4q)} \sim  1.0 \times 10^{-5}$ at the end of the hadronic phase.

On the other hand, if the $X(3872)$ is a molecular state it will be produced by hadron coalescence at the
end of the hadronic phase. According to Ref.~\cite{ChoLee}, at this time 
the average number of $X$'s, considering it as a  $D\bar{D}^*$ molecule, is $N_{X(mol)} \approx  7.8 \times 10^{-4}$, which is about $80$ times larger than $N_{X(4q)}$.

As expected, the results show that  the $X$ multiplicity in relativistic ion collisions  depends on the structure of $X(3872)$. Our main conclusion is 
that  the  contribution from the anomalous vertices play an important role in determining the time evolution of the $X(3872)$ abundance and  
they lead to strong suppression of this state during the hadronic phase. Therefore, within the uncertainties of our 
calculation we can say that if the $X(3872)$ were observed in a heavy ion collision it must have
been produced at the end of the hadronic phase and, hence, it must be a molecular state.

\begin{acknowledgements}

The authors would like to thank the Brazilian funding agencies CNPq and FAPESP for financial support. We also thank C. Greiner and J. Noronha-Hostler for fruitful 
discussions.

\end{acknowledgements}


\end{document}